# Non-excitonic mechanism for electronic and structural phase transitions in $Ta_2Ni(Se,S)_5$

Weichen Tang[1,2], Zhenglu Li[1,2,3], Cheng Chen[4,5], Yu He[5], Steven G. Louie[1,2]†

[1]*Physics Department, University of California, Berkeley, California 94720, USA*

[2]*Materials Sciences Division, Lawrence Berkeley National Lab, Berkeley, California 94720, USA*

[3]*Mork Family Department of Chemical Engineering and Materials Science, University of Southern California, Los Angeles, California 90089, USA*

[4]*Department of Physics, University of Oxford, Oxford, OX1 3PU, United Kingdom*

[5]*Department of Applied Physics, Yale University, New Haven, Connecticut 06511, USA*

†sglouie@berkeley.edu

Abstract: We present a first-principles study based on density functional theory (DFT) on the electronic and structural properties of $Ta_2NiSe_5$, a layered transition metal chalcogenide that has been considered as a possible candidate for an excitonic insulator. Our systematic DFT results however provide a non-excitonic mechanism for the experimentally observed electronic and structural phase transitions in $Ta_2NiSe_5$, in particular explaining why sulfur substitution of selenium reduces the distortion angle in the low-temperature phase and potassium dosing closes the gap in the electronic structure. Moreover, the calculations show that these two effects couple to each other. Further, our first-principles calculations predict several changes in both the crystal structure and electronic structure under the effects of uniform charge dosing and uniaxial strain, which could be tested experimentally.

The study of the excitonic insulator phase of matter represents an exciting, emerging frontier in condensed matter physics. This correlated phase of matter is characterized by the spontaneous formation of excitons, which are bound electron-hole pairs arising from their Coulomb interaction. If the binding energy of the excitons ($E_b$) exceeds the material's quasiparticle band gap ($E_g$), the system may undergo a transition from its normal band insulator ground state into an excitonic insulator state [1,2] below certain critical temperature. Despite the seemingly straightforward concept and conditions required for its formation, the definitive determination of whether a material is an excitonic insulator is challenging. One key challenge stems from the intricate interplay and possibility of multiple degrees of freedom that collectively contribute to the phase transition [3]. For instance, in the stoichiometric indirect band gap semiconductor 1T-$TiSe_2$ [4,5], both plasmons and phonons exhibit a softening to zero energy below the phase transition temperature. This observation hints at a combined effect from both the lattice and electronic degrees of freedom, and recent studies further challenged the role of an exciton condensate behind the phase transition [6,7].

Another material that has generated substantial debate as a potential excitonic insulator candidate is $Ta_2NiSe_5$, a layered transition metal chalcogenide material. This compound has garnered significant attention due to its intriguing structural and electronic properties akin to the expectation of an excitonic insulator. At around 328 K, $Ta_2NiSe_5$ undergoes a structural phase transition from a high-temperature orthorhombic phase to a low-temperature monoclinic phase, without the formation of superstructures. This transition is accompanied by a change from a metallic to an insulating state. Several experimental observations – such as the camelback-shaped valence band top in angle-resolved photoemission spectroscopy (ARPES) experiments, the electronic structure change upon surface potassium dosing, and an anomalous absence of zone center phonon softening – can be rationalized within the framework of transition to an excitonic insulator phase [8–18]. However, recent studies suggest that lattice distortions also play a crucial role in facilitating the electronic phase transition, while the contribution of excitonic effects may be parasitic [19–23].

When multiple degrees of freedom come into play, first-principles investigations can often shed light on a phenomenon, as they do not rely significantly on empirical or adjustable parameters. However, in the case of $Ta_2NiSe_5$, previous density functional theory (DFT) studies performed with different exchange-correlation functionals yielded conflicting results. Two primary issues emerged: one centers on whether the DFT ground state corresponds to a monoclinic or orthorhombic structure, while the other one focuses on whether a subtle distortion angle can produce a significant energy gap in line with experimental observations [13–15,19,21,24]. The persistent controversy on whether $Ta_2NiSe_5$ is an excitonic insulator at low temperature can be partially attributed to the diverse results and analyses arising from different DFT calculations. Each of these analyses and their corresponding conclusions are consistent with a subset of experimental results, making it difficult to be definitive when considering $Ta_2NiSe_5$ in isolation.

A more robust strategy to resolve the appropriateness of a functional choice is to subject the investigation to the test of continuously modified compounds – in this case, the isovalent substitution series $Ta_2Ni(Se,S)_5$. Recent experiments have unveiled intriguing phenomena by introducing sulfur doping (in place of selenium), enabling precise control over the phase transition [25–28]. By systematically varying the sulfur content, one can observe how the structural and electronic properties evolve, providing a more comprehensive test for the accuracy and reliability of different DFT exchange-correlation functionals in capturing the complex interplay between the system's crystal structure and electronic behavior. In this way, we gain the ability to assess the performance of different DFT functionals across the entire series. As the level of sulfur doping varies continuously in the system, it becomes possible to identify an optimal functional that best describes the $Ta_2Ni(Se,S)_5$ material class.

Following this strategy, we have performed systematic DFT calculations on $Ta_2Ni(Se,S)_5$ and our results suggest a non-excitonic mechanism for the electronic and structural phase transitions in $Ta_2NiSe_5$. Based on the criterion of selecting a functional that accurately captures both the electronic and structural properties across the entire series for $Ta_2Ni(Se,S)_5$, we arrive upon the $r^2$SCAN functional [29] as being the most appropriate functional for this class of materials. The structure relaxation calculations yield a low-temperature monoclinic crystal structure with a small distortion angle of 0.49°, agreeing with the experimental value of 0.47°-0.67°. We further investigate the effects of sulfur doping and potassium dosing on the phase transition and the electronic structure of $Ta_2NiSe_5$. We find that replacing selenium with sulfur suppresses the structural phase transition at low temperature, which is ultimately eliminated in $Ta_2NiS_5$ (100% replacement). Furthermore, we observe that uniform electron doping in both $Ta_2NiSe_5$ and $Ta_2NiS_5$ in the low-temperature phase eliminates the band gap. These alloying and doping results are consistent with experimental observations. However, our DFT study reveals that the mechanisms leading to the electronic structure transitions are different for these two approaches. Finally, we make predictions about lattice structure changes with potassium dosing and the effects of strain on the system that could be experimentally verified. Our findings thus capture the experimental results without needing additional (e.g., BCS-type) treatment with electron-hole Coulomb interactions for a correlated phase.

Figure 1(a) displays the crystal structure of $Ta_2NiSe_5$. The high-temperature structure is orthorhombic (*Cmcm*), while the low-temperature structure is monoclinic (*C2/c*) which has lower symmetry. The DFT structure relaxation results using $r^2$SCAN yield a monoclinic structure with lattice parameters $a = 3.52$Å, $b = 12.63$Å, $c = 15.51$Å, $\beta = 90.49°$, agreeing with the experimental values $a = 3.50$Å, $b = 12.87$Å, $c = 15.68$Å, $\beta = 90.67°$. The distortion angle defined as $(\beta - 90°)$ in the low-temperature phase is used as a parameter to indicate the strength of the structural transition. To qualitatively explore the thermal effect on electrons, we use a Fermi-Dirac smearing to estimate the temperature effect as explained in the Supplemental Material [30]. This approach neglects phonon contributions, which may also play a significant role. Nevertheless, we find that an increase in the effective temperature for the electrons resulted in a decrease in the $\beta$ angle, consistent with the observed monoclinic to orthorhombic transition at high temperature.

Sulfur doping (with sulfur replacing selenium) is known for both reducing the ground state structural distortion and tuning the electronic structure of the system [25–28]. We first examine the effect of S doping on the crystal structure. The change in $\beta$ as a function of S doping level $x$ are presented in Fig. 1(b). The results show that $\beta$ decreases as the S doping level $x$ increases. With the system eventually transforms to $Ta_2NiS_5$ (correspond to $x = 1$), $\beta$ approaches 90°, indicating the absence of a structural phase transition in this case. Our DFT structure relaxation results agree with recent high-resolution synchrotron X-ray diffraction (XRD) results [28], suggesting that substituting Se with S weakens the phase transition in the system. Additionally, in the Supplemental Material [30], the temperature effect at various S doping levels is estimated, qualitatively showing a decrease in the phase transition temperature with increasing S doping level, while noting that phonon effects are not included in this analysis.

The valence band top of the monoclinic phase of $Ta_2NiSe_5$ exhibits a camelback feature observed experimentally, which was previously interpreted to be associated with the excitonic insulator phase [8]. We show that this is captured by our DFT electronic structure calculations without invoking explicitly any long-range non-Hartree interactions. Although the DFT Kohn-Sham bandgap is smaller than the experimental value as expected, which is primarily due to the neglect of self-energy effects, this discrepancy does not affect our main conclusions regarding the electronic ground-state properties and the physical origin of the metal-insulator transition. Furthermore, our previous work [28] demonstrated that the high-symmetry phase of $Ta_2Ni(Se,S)_5$ has a metal-to-insulator transition along the S doping axis (around $x$=0.7), while the ground state is always an insulator. This finding resolves the inconsistency between previous resistivity measurements (which indicate a phase boundary $x_c$=0.7) [25] and the XRD results, where the lack of apparent resistive anomaly above $x_c$ is due to a readily gapped normal state.

We now consider the effect of surface potassium dosing (i.e., introducing excess electron charge) on the ground state. Previous research suggested that potassium dosing could restore the electronic structure to its normal phase (the high-temperature phase) [12,16,23]. In the case of $Ta_2NiSe_5$, the electronic structure undergoes an insulator-to-metal transition upon potassium dosing, which was previously attributed to the suppression of the excitonic insulator phase [12,16]. However, it remains unclear whether potassium dosing induces a change in the crystal structure. If so, the observed insulator-to-metal transition could merely be a concurrent effect resulting from the structural change.

We present DFT-calculated electronic structures of $Ta_2Ni(Se,S)_5$ in the lowest-energy configuration with potassium dosing ($0.4e^-$/unit cell) and without dosing in Fig. 2, showing excellent agreement with ARPES data. Regardless of the S content, the system at T = 0 undergoes a semiconductor to semimetal transition with an injection of $0.4e^-$/unit cell. Notably, for high S content (corresponding to $x > x_c$), the high-symmetry phase is a semiconductor. This finding implies that the previous understanding that potassium dosing could restore the system to its high-symmetry phase is only valid for low S content (corresponding to $x < x_c$).

Since $Ta_2NiS_5$ is a normal semiconductor at T = 0, the insulator-to-metal transition under potassium dosing cannot be captured within the conventional framework of an excitonic insulator transition. A different mechanism to account for the electronic structure transition of $Ta_2Ni(Se,S)_5$ under potassium dosing is explored below.

We first examine the crystal structure during the charge dosing process. Fig. 3(a) and (b) illustrates the changes in lattice constants and the distortion angle $\beta$ under different conditions. DFT results indicate that when the S content is low (x close to 0), not only do the lattice constants change, the symmetry of the system also changes. Specifically, upon sufficient charge dosing, a crystal structure transition from monoclinic to orthorhombic (with $\beta$ decreasing to 90°) takes place.

To confirm that the crystal symmetry transition plays a dominant role in the electronic structure transition, we plot the $\Gamma - \Gamma$ gap/overlap (defined as the energy difference between the lowest conduction band and the highest valence band at the Γ point) as a function of potassium dosing level in Fig. 3(c). We consider two cases: 1) allowing both the symmetry of the system (i.e., changing of the angle $\beta$ from 90°) and the lattice constants to change, and 2) only allowing the lattice constants to change. DFT calculations suggest that the change of $\beta$ from 90° dominates the changes in the electronic structure. When both $\beta$ and lattice constants are allowed to change, the $\Gamma - \Gamma$ gap changes from 91 meV to -171 meV from 0 to $0.4e^-$/unit cell. In contrast, when $\beta$ is fixed at 90° and only the lattice constants are allowed to change, the $\Gamma - \Gamma$ gap changes from a -75 meV overlap to -171 meV overlap in the same range of dosing. This result indicates that the dosing-induced change in symmetry has a more substantial impact on the electronic structure than merely changes in the lattice constants.

However, when the S content is high (x close to 1), the deviation of the angle $\beta$ from 90° is small, especially in the case of $Ta_2NiS_5$ where $\beta$ remains at 90°. Fig. 3(d) shows that if we maintain the lattice constants unchanged, the $\Gamma - \Gamma$ gap/overlap remains almost the same. However, if we allow all lattice parameters to change, the $\Gamma - \Gamma$ gap/overlap changes from 131 meV to -10 meV. This result suggests that, at full sulfur substitution, the change of lattice constants dominates the $\Gamma - \Gamma$ gap evolution with potassium dosing.

In both scenarios, potassium dosing alters the lattice structure, which subsequently leads to a band overlap at sufficient dosing. It is thus intriguing to investigate the possibility of controlling electronic properties via other means of structure modifications. Applying strain is a method to precisely control the lattice constant in a desired direction. Given that the lattice constant *b* is most sensitive to potassium dosing (Fig. 3(a)), we have considered applying strain in the out-of-plane direction (perpendicular to the *a*-*c* plane of distortion).

Fig. 4 presents our predictions regarding the impact of strain on both the crystal structure and electronic structure of $Ta_2NiSe_5$. As depicted in Fig. 4(a), when positive strain (extension) is applied parallel to the *b*-axis, the $\beta$ angle remains nearly constant. However, applying negative strain (compression) along the *b*-axis results in a progressive decrease in the $\beta$ angle until it reaches 90 degrees at -8% strain. One might hypothesize that the change in the distortion in the angle $\beta$ would lead to an insulator-to-metal transition, similar to the case under potassium dosing. However, since we directly modify the lattice constant *b*, the electronic structure changes are actually more intricate.

Fig. 4(c) and (d) illustrate the electronic structure of $Ta_2NiSe_5$ under +4% and -4% strain, respectively, which are to be compared to the electronic structure at equilibrium in Fig. 4(b). With positive strain applied, although the $\beta$ angle remains nearly unchanged (see Fig. 4(a)), the band gap near the Γ point closes with a slight overlap between the conduction and valence bands, and the electronic structure undergoes a transition from insulator to metal. In contrast, under negative strain at -4%, despite the restoration towards the orthorhombic structure, the band gap near the Γ point actually increases. Additionally, along the *k* direction from the Γ point to the X point, bands that were originally further below the Fermi level rise substantially, creating a much easier condition to gate the system towards a multi-band Fermi surface. These predictions can be readily verified with experiments.

The pronounced sensitivity of the electronic structure to lattice distortions in $Ta_2Ni(Se,S)_5$ suggests that electron-phonon coupling may play a more central role than previously appreciated. Such coupling, along with possible anharmonic effects, could contribute to phonon renormalization and linewidth broadening. The order-parameter-like suppression of the 1 THz phonon amplitude [11] might reflect enhanced electronic dressing near the transition [23] or nonlinear phonon interactions, while the lack of phonon softening across phase transition [18] could stem from the interplay between strong electron-phonon coupling and gap opening. These scenarios, though still speculative, raise the possibility that key phonon anomalies may also be explained without invoking excitonic effects. A systematic study of phonon dynamics across the $Ta_2Ni(Se,S)_5$ series can help clarify these contributions and is reserved for future work [36].

In conclusion, we utilize first-principles DFT calculations to investigate the electronic and structural phase transitions in the $Ta_2NiSe_5$ system. The DFT results are validated by XRD and ARPES data for various S doping and potassium dosing levels. We demonstrate that the electronic structure of the system undergoes a transition from insulator to metal upon potassium dosing for any S doping level, as supported by both theoretical and experimental evidence. Moreover, we clarify that the main contributor to the electronic structure transition differs for small and large S doping levels, and the predicted changes in lattice structure during potassium dosing can be confirmed through future experiments. Lastly, we offer predictions on the effects of out-of-plane strain on both the crystal structure and electronic structure of the system. Unless non-local electron–hole interactions, which are needed for the formation of bound excitons, are inadvertently captured by the semi-local functional employed in our study, our findings provide an alternative non-excitonic mechanism for the phase transition observed in the $Ta_2Ni(Se,S)_5$ family.

This work was primarily supported by the Theory of Materials Program (No. KC2301) funded by the U.S. Department of Energy, Office of Basic Energy Sciences under Contract No. DE-AC02-05CH11231 at the Lawrence Berkeley National Laboratory. Computational resources were provided by the National Energy Research Scientific Computing Center (NERSC), which is supported by the Office of Science of the U.S. Department of Energy under Contract No. DEAC02-05CH11231, Stampede3 at the Texas Advanced Computing Center (TACC), through Advanced Cyberinfrastructure Coordination Ecosystem: Services & Support (ACCESS), which is supported by National Science Foundation under Grant No. ACI1053575, and Frontera at TACC, which is supported by the National Science Foundation under Grant No. OAC1818253. The photoemission work is supported by National Science Foundation (NSF) under DMR-2239171. Use of the Stanford Synchrotron Radiation Light Source, SLAC National Accelerator Laboratory, is supported by the US Department of Energy, Office of Science, Office of Basic Energy Sciences under Contract No. DE-AC02-76SF00515. The authors thank Z. Kang, and Y. Wang for helpful discussions, and X. Chen and R. Birgeneau for providing the samples.

The data that support the findings of this article are openly available [37], embargo periods may apply.

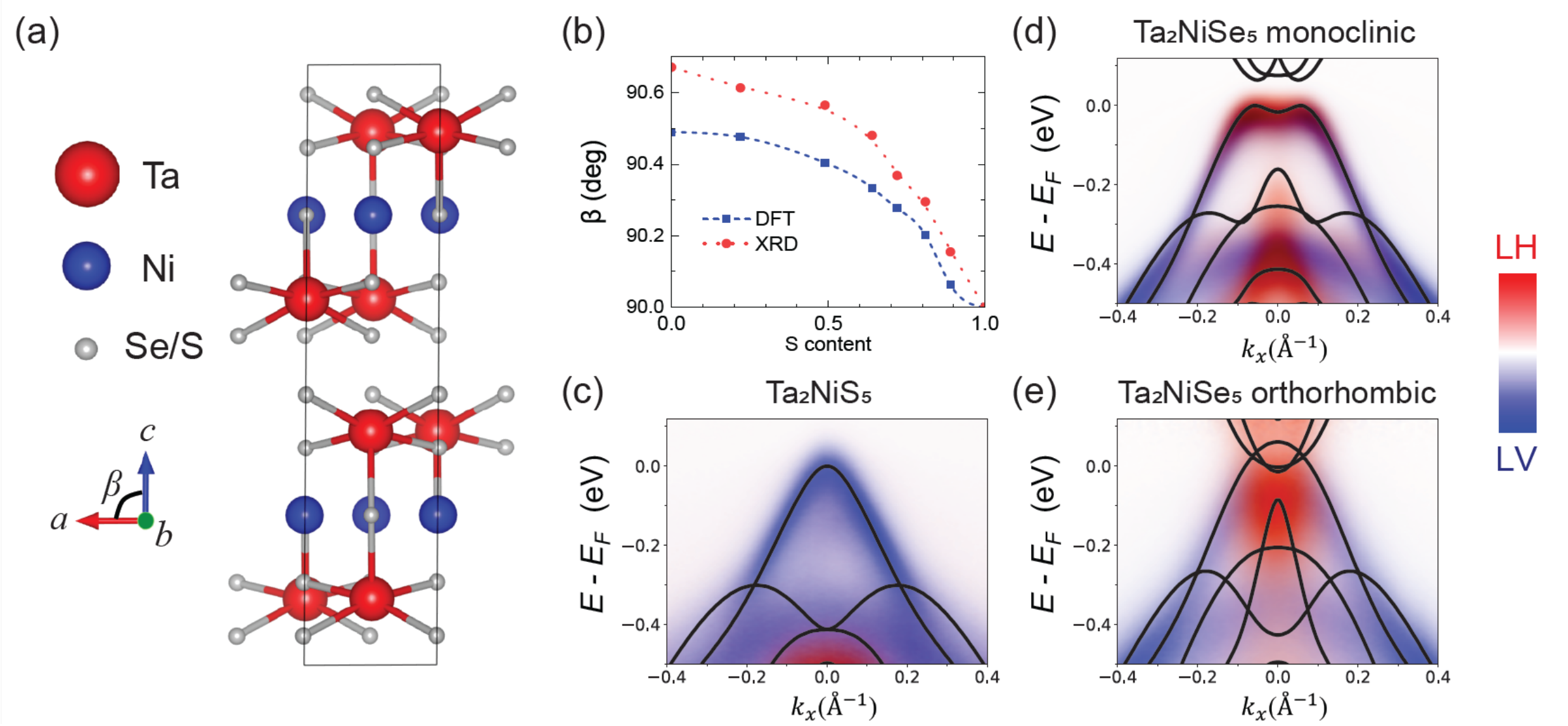

Figure 1 (a) Crystal structure of $Ta_2Ni(Se,S)_5$. The $\beta$ angle is defined as the angle between the lattice vector $a$ and lattice vector $c$. In the orthorhombic phase of $Ta_2Ni(Se,S)_5$, $\beta = 90°$, while in the monoclinic phase, $\beta$ slightly deviates from 90°. (b) Comparison between DFT calculations and experimental XRD results on the change of $\beta$ with S doping level. (c)&(d) Comparison between DFT electronic structures and ARPES data for $Ta_2NiS_5$ and $Ta_2NiSe_5$: The ARPES spectra are displayed using a 2D color map, where the red color indicates the linear horizontal (LH) polarization of incident light and the blue color indicates the linear vertical (LV) polarization. The DFT results are represented by solid lines. The reciprocal lattice vectors are aligned along the X-Γ-X line, which runs parallel to the direction of the lattice vector $a$.

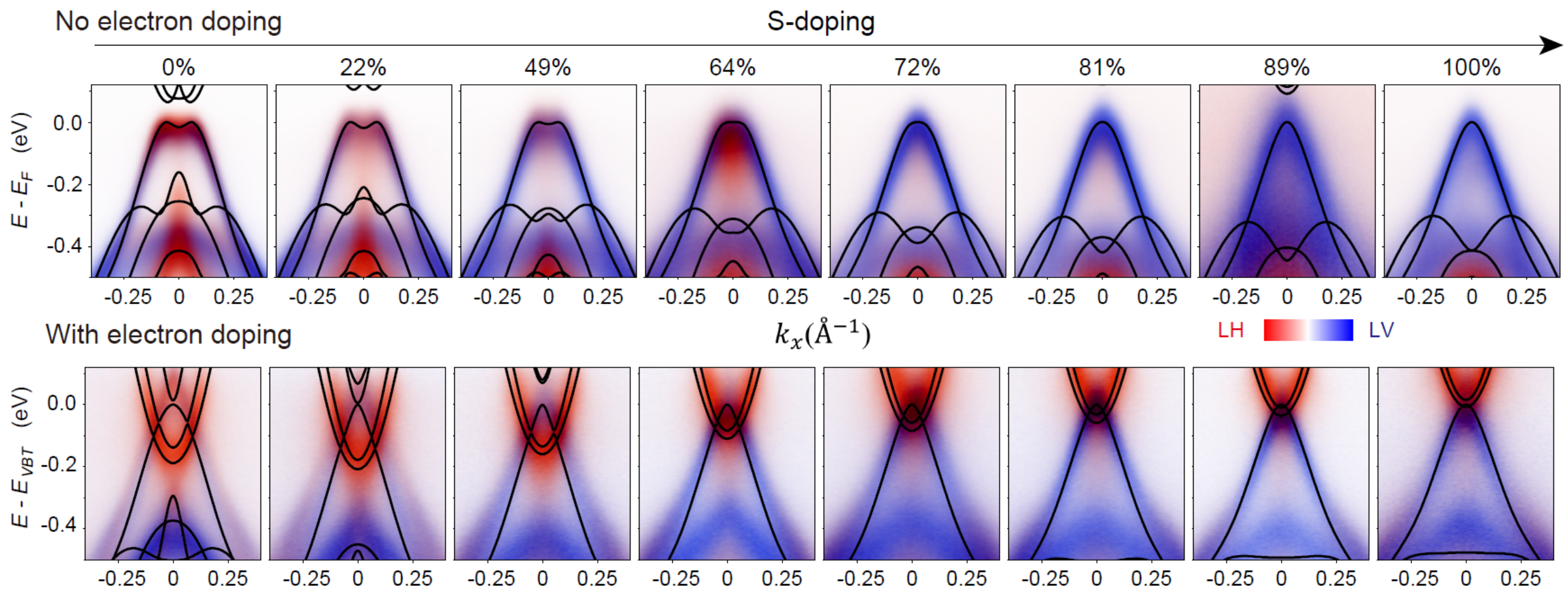

Figure 2 Comparison between DFT calculations and experimental APRES data of electronic structure of $Ta_2Ni(Se,S)_5$ without (top panels) and with (bottom panels) $0.4e^-$/unit cell potassium dosing. The ARPES spectra are displayed using a 2D color map and the DFT results are represented by solid lines. In lower panels for results upon K dosing, the energy is aligned with respect to the valence band top (VBT).

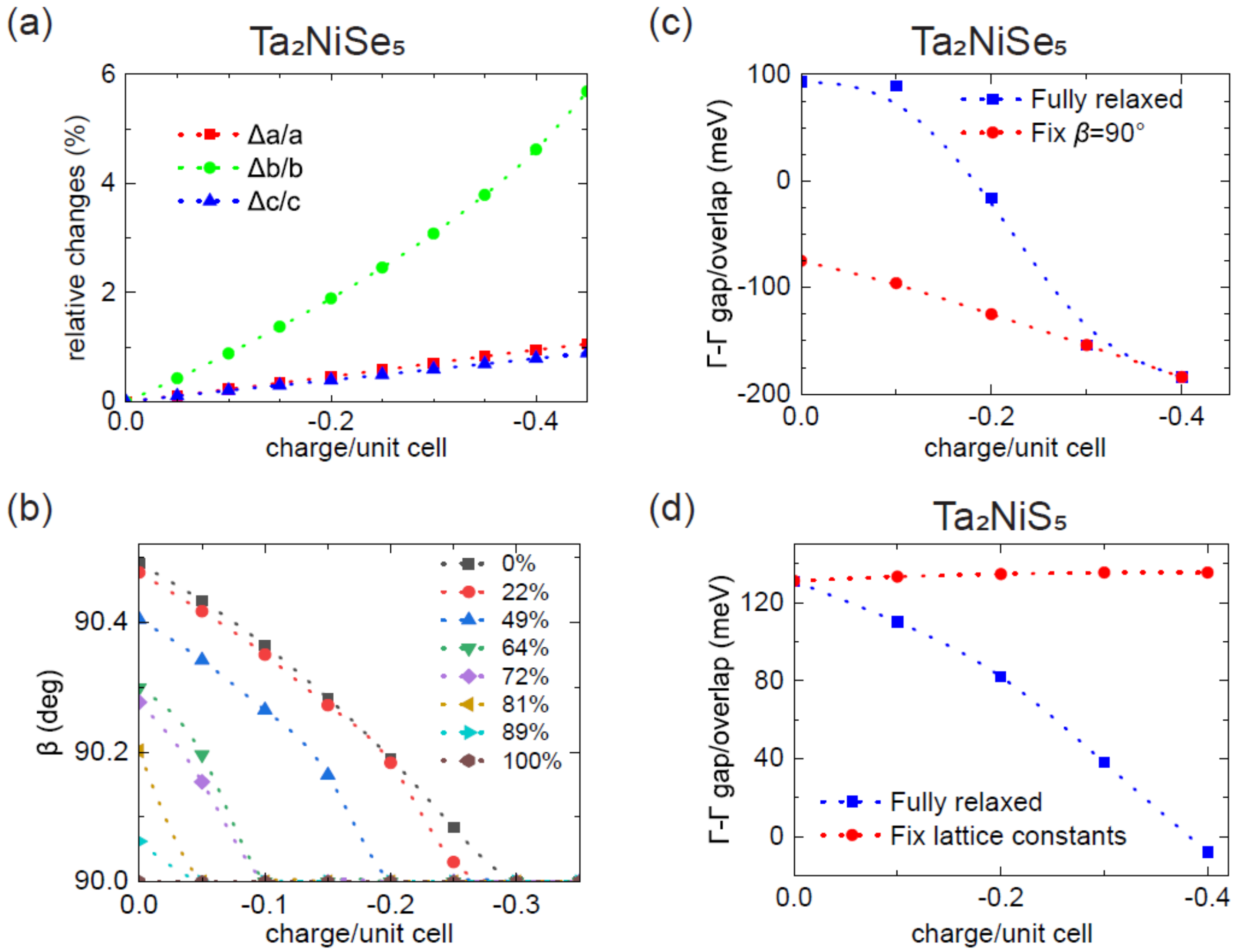

Figure 3 (a) The change in lattice constants with the introduction of extra electrons in $Ta_2NiSe_5$. Similar behavior exists for other sulfur contents. (b) The change in distortion angle $\beta$ with the introduction of extra electrons for different S doping levels given in percentage. (c)&(d) The $\Gamma-\Gamma$ gap/overlap as a function of the level of potassium dosing for $Ta_2NiSe_5$ and $Ta_2NiS_5$, respectively. In (c) for $Ta_2NiSe_5$, we compare results with fixing $\beta = 90°$ against with allowing all lattice parameters to change. In (d) for $Ta_2NiS_5$, we compare results with fixing the lattice constants against with allowing all lattice parameters to change.

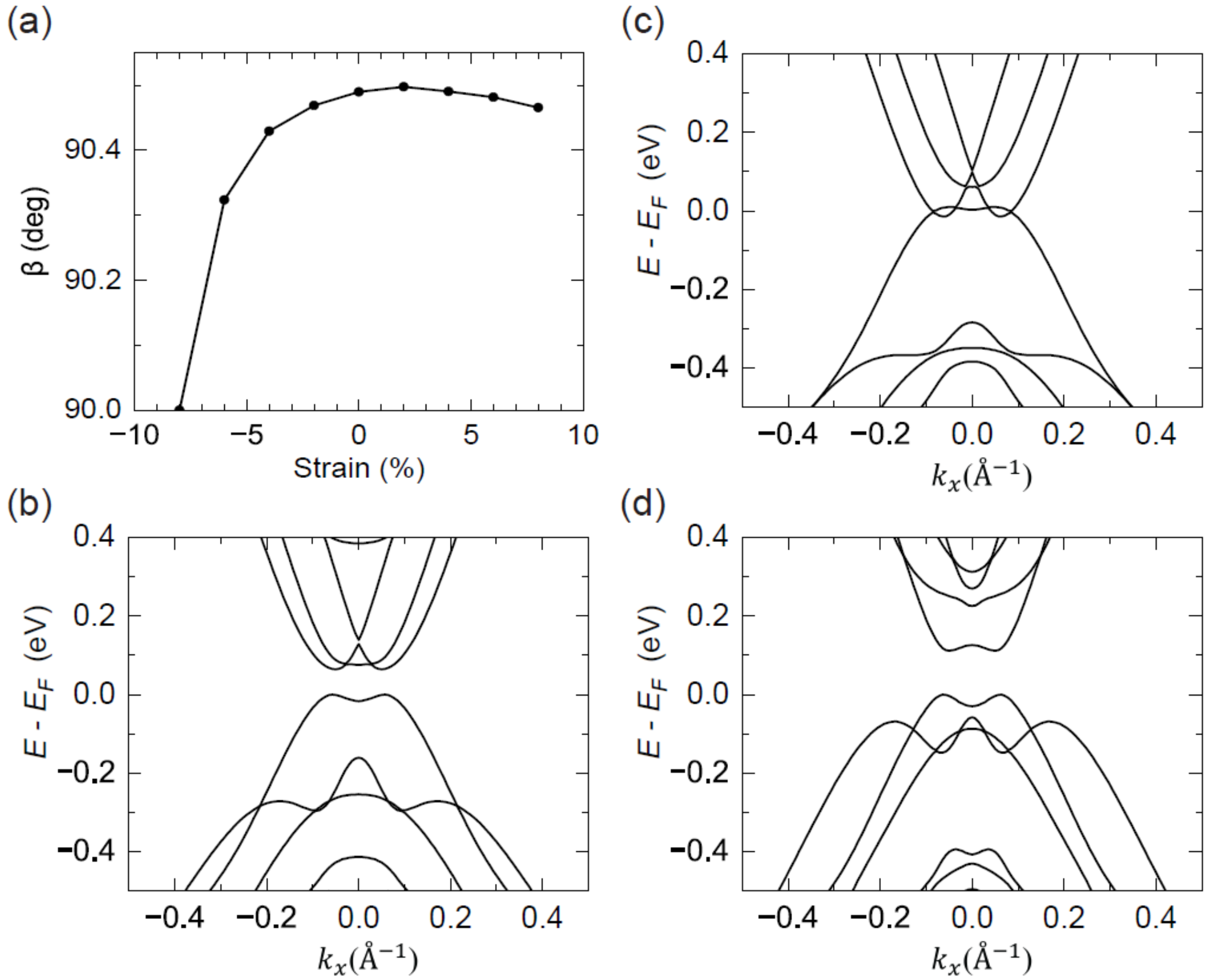

Figure 4 The change of crystal structure and electronic structure of $Ta_2NiSe_5$ under strain along the direction of the *b* lattice vector. (a) Change of the distortion $\beta$ angle as a function of strain. (b) DFT calculated band structure of $Ta_2NiSe_5$ at equilibrium. (c) DFT calculated band structure of $Ta_2NiSe_5$ under +4% strain along the direction of the *b* lattice vector. (d) DFT calculated band structure of $Ta_2NiSe_5$ under -4% strain along the direction of the *b* lattice vector. The **k** vectors are aligned along the X-Γ-X line, which runs parallel to the direction of the *a* lattice vector.